\documentclass[
aps,
nofootinbib,
prd,
superscriptaddress,
tightenlines,
notitlepage,
twocolumn,
showpacs
]{revtex4-1}

\usepackage[colorlinks]{hyperref}
\usepackage{tabularx}
\usepackage{graphicx}
\usepackage{amssymb,amsmath,bm,tensor,braket}
\usepackage{xcolor}
\usepackage[varg]{txfonts}
\usepackage{enumerate}
\usepackage{mathtools}
\usepackage[capitalize]{cleveref}
\crefname{section}{Sec.}{Secs.}
\Crefname{section}{Section}{Sections}
\crefname{equation}{Eq.}{Eqs.}
\Crefname{equation}{Equation}{Equations}
\crefname{figure}{Fig.}{Figs.}
\Crefname{figure}{Figure}{Figures}
\crefname{subsection}{Sec.}{Secs.}
\usepackage[utf8]{inputenc}
\usepackage[normalem]{ulem}

\newcommand*\rmd{\ensuremath{\mathop{}\!\mathrm{d}}}
\newcommand*\ii{\ensuremath{\mathop{}\!\mathrm{i}}}
\newcommand*\DL{\ensuremath{D_{\rm L}}}
\newcommand*\calA{\ensuremath{\mathcal{A}}}
\newcommand*\calM{\ensuremath{\mathcal{M}}}

\newcommand*\htilde{\ensuremath{\widetilde{h}}}

\newcommand{\BNU}{\affiliation{Department of Astronomy, Beijing Normal
University, Beijing 100875, China}}
\newcommand{\DOA}{\affiliation{Department of Astronomy, School of Physics,
Peking University, Beijing 100871, China} }
\newcommand{\KIAA}{\affiliation{Kavli Institute for Astronomy and
Astrophysics, Peking University, Beijing 100871, China}}
\newcommand{\SOP}{\affiliation{School of Physics and State Key Laboratory
of Nuclear Physics and Technology, Peking University, Beijing 100871,
China}}
\newcommand{\CICQM}{\affiliation{Collaborative Innovation Center of Quantum
Matter, Beijing, China}}
\newcommand{\CHEP}{\affiliation{Center for High Energy Physics, Peking
University, Beijing 100871, China}}
\newcommand{\NAOC}{\affiliation{National Astronomical Observatories,
Chinese Academy of Sciences, Beijing 100012, China}}

\begin{document}

\title{Probing dipole radiation from binary neutron stars with ground-based
laser-interferometer and atom-interferometer gravitational-wave
observatories}

\date{\today}
\author{Junjie Zhao}\SOP
\author{Lijing Shao}\email[Corresponding author: ]{lshao@pku.edu.cn}\KIAA\NAOC
\author{Yong Gao}\DOA\KIAA
\author{Chang Liu}\DOA\KIAA
\author{Zhoujian Cao}\BNU
\author{Bo-Qiang Ma}\SOP\CICQM\CHEP

\begin{abstract} 

Atom-interferometer gravitational-wave (GW) observatory, as a new design of
ground-based GW detector for the near future, is sensitive at a relatively
low frequency for GW observations. Taking the proposed atom interferometer
Zhaoshan Long-baseline Atom Interferometer Gravitation Antenna (ZAIGA), and
its illustrative upgrade (Z+) as examples, we investigate how the atom
interferometer will complement ground-based laser interferometers in
testing the gravitational dipole radiation from binary neutron star (BNS)
mergers. A test of such kind is important for a better understanding of the
strong equivalence principle laying at the heart of Einstein's general
relativity. To obtain a statistically sound result, we sample BNS systems
according to their merger rate and population, from which we study the
expected bounds on the parameterized dipole radiation parameter $B$.
Extracting BNS parameters and the dipole radiation from the combination of
ground-based laser interferometers and the atom-interferometer ZAIGA/Z+, we
are entitled to obtain tighter bounds on $B$ by a few times to a few orders
of magnitude, compared to ground-based laser interferometers alone,
ultimately reaching the levels of $|B| \lesssim 10^{-9}$ (with ZAIGA) and $|B|
\lesssim 10^{-10}$ (with Z+).

\end{abstract}

\maketitle

\section{Introduction}
\label{sec:intro}

Einstein's theory of general relativity (GR) has been tested for more than
a century via a great number of accurate experiments~\cite{Will:2018bme}.
In GR gravitation is described only by a spin-2 tensor field, $g_{\mu
\nu}$~\cite{Einstein:1915ca}, with the corresponding gravitational emission
from compact binary coalescences dominated by the quadrupolar
radiation~\cite{Einstein:1916cc, Einstein:1918btx}. For the radiative tests
in GR, the Hulse-Taylor binary pulsar offered the first empirical evidence
for the existence of gravitational waves (GWs)~\cite{Hulse:1974eb}. The
first direct detection at the Earth of a GW signal, GW150914, has been
announced by the Laser Interferometer GW Observatory (LIGO) Scientific
Collaboration and Virgo Collaboration~\cite{Abbott:2016blz}. Since then,
the number of GW events has expanded significantly in the past five
years~\cite{LIGOScientific:2018mvr, LIGOScientific:2020ibl} and they have
stimulated fruitful researches concerning our gravitational Universe.

Although GR passes all the existing tests with flying colors, there remain
some well-motivated alternative gravity theories, including one or more
extra degrees of freedom other than $g_{\mu \nu}$ in the gravitational
sector. These theories have a demonstrated potential to solve problems in
the inflation of the Universe, dark energy, as well as to serve as a
prototype for a unified theory of quantum gravity~\cite{Fujii:2003pa}.
Different from GR, in some of these alternative theories, dipole
gravitational radiation from coalescing binaries is no longer strictly
forbidden. Meanwhile, there are good theoretical supports that a nonzero
dipole radiation violates the strong equivalence principle (SEP), which is
at the heart of GR~\cite{Will:2014kxa,Shao:2016ezh, Barausse:2017gip}.
Therefore, in principle, searching for a dipole radiation directly tests
the SEP~\cite{Barausse:2017gip}. As a result, some aspects of strong-field
gravity closely related to the SEP can be probed from the dipole radiation
tests in binary pulsars and compact binary coalescences.

In the quasi-stationary strong-field regime, binary pulsar systems are
powerful testbeds for gravity, where at least one component of the binary
is strongly self-gravitating and the characteristic orbital speed of the
binary is small while compared with the light speed~\cite{Wex:2014nva}. Due
to the extremely precise technology, the so-called {\it pulsar timing}, the
existence of dipole radiation has been excluded to high confidence in these
systems~\cite{Freire:2012mg, Yagi:2013ava, Shao:2017gwu, Zhao:2019suc, Guo:2021leu}.

In recent years, the advanced LIGO (AdvLIGO) and the advanced Virgo (AdV)
together have discovered a bulk of GW events~\cite{LIGOScientific:2018mvr,
LIGOScientific:2020ibl}, which sparked great enthusiasm in fields of gravity
tests~\cite{TheLIGOScientific:2016src, Abbott:2018lct, LIGOScientific:2020tif,
GBM:2017lvd, Shao:2020shv, Akbarieh:2021vhv, Sathyaprakash:2019yqt}. Those discoveries make it possible to investigate various
properties of gravitation in highly-dynamical, strong-field, as well as
radiative regimes~\cite{Wex:2014nva,Barausse:2016eii}. GWs have become a
new testbed for testing strong-field gravity, in addition to the binary
pulsar systems~\cite{Damour:2007uf}. For some alternative theories, the
additional gravitational degrees of freedom not only change the orbital
binding energy of binaries, but also contribute to the GW waveform
significantly in such strong-field regimes~\cite{Damour:1998jk,
Shao:2017gwu}. Therefore, GWs with the dipole-radiation imprints could be
detected or constrained via GW observations.

The dipole-radiation effect, corresponding to the $-1$ post-Newtonian (PN)
contribution, enters the GW waveform at a lower order than the canonical
quadrupolar radiation in GR.\footnote{The correction of $n$\,PN refers to
$\mathcal{O}\left((v/c)^{2n} \right)$ corrections with respect to the
Newtonian order, where $v$ is the characteristic relative speed in the
binary. Correspondingly, the quadrupolar radiation is $0$\,PN contribution
in this convention.} Therefore, the dipole radiation is relatively more
significant at a lower GW frequency. In principle, the GW observatories
with a lower accessible frequency could obtain more GW information about
negative PN corrections. Such reasoning applies to the dipole radiation in
this work.

Until now, LIGO Scientific Collaboration and Virgo Collaboration (LVC) have
officially announced 50 compact binary coalescences in total from observing
runs O1, O2, and O3a~\cite{LIGOScientific:2018mvr,LIGOScientific:2020ibl}.
Recently, LVC is upgrading their detectors and planning to launch the
observing run O4 in late 2021 or early 2022~\cite{Abbott:2020qfu}. In
addition, the Kamioka GW Detector
(KAGRA)~\cite{Akutsu:2018axf,Akutsu:2020zlw}, which is built in Japan,
joined O3 on October 4, 2019, and the four-detector network has officially
begun~\cite{Abbott:2020qfu}. Besides, more future GW detectors are proposed
worldwide, including
LIGO-India,\footnote{\url{https://dcc.ligo.org/LIGO-M1100296/public}} and
the third-generation GW detectors, the Cosmic Explorer (CE) led by the
United States~\cite{Reitze:2019iox,Reitze:2019dyk} and the Einstein
Telescope (ET) led by Europe~\cite{Hild:2010id, Sathyaprakash:2019yqt}.

Over the coming decades, the ground-based detectors, AdvLIGO, AdV, and KAGRA
will upgrade their instruments further, and eventually, the
third-generation detectors will be online. By that time, hundreds of
thousands of stellar-mass binaries can be discovered across the GW
frequency band in $10$--$10^{3} \, {\rm Hz}$~\cite{Reitze:2019dyk}. In the
2030s, the space-borne detectors like Laser Interferometer Space Antenna
(LISA)~\cite{Audley:2017drz}, Taiji~\cite{Luo:2020taiji} and
TianQin~\cite{Mei:2020lrl} will enable the GW observations of the massive
black holes (BHs) in the range of GW frequency $10^{-4}$--$10^{-1} \, {\rm
Hz}$. Benefiting from larger interferometer arms and quieter space
environment, the space-borne detectors are suitable to investigate the
low-frequency GW sciences by design. \citet{Sesana:2016ljz} studied the
prospects for multi-band GW observations with AdvLIGO and LISA. It is shown
that LISA can inform AdvLIGO in advance when and where binary BH (BBH)
coalescences will happen. \citet{Barausse:2016eii} used the multi-band
observations of massive BBHs in AdvLIGO and LISA to predict the projected
bounds on dipole radiation.

Given the above detectors, there remains a frequency gap spanning $10^{-1}
$--$ 10\,{\rm Hz}$. To fulfill it, a bulk of decihertz GW detectors are
proposed, broadly speaking including (i) space-borne laser interferometer
detectors like the Decihertz Observatory (DO)~\cite{Sedda:2019uro, Sedda:2021hpg} and the
DECihertz laser Interferometer GW Observatory
(DECIGO)~\cite{Sato:2017dkf,Kawamura:2020pcg}, and (ii) ground-based atom
interferometer detectors. Recently, \citet{Liu:2020nwz} investigated the
projected constraints on dipole radiation from AdvLIGO, LISA and the
space-borne decihertz detectors. Similar tests of GR via the multi-band
observations involving decihertz detectors were also investigated in
Refs.~\cite{Gnocchi:2019jzp,Carson:2019rda}. Concerning decihertz GW
detectors on the ground, atom interferometers---such as Zhaoshan
Long-baseline Atom Interferometer Gravitation Antenna
(ZAIGA)~\cite{Zhan:2019quq}, Matter wave-laser based Interferometer
Gravitation Antenna (MIGA)~\cite{Canuel:2017rrp}, and a next-generation
Atomic Interferometric Observatory and Network
(AION)~\cite{Badurina:2019hst}---are proposed to investigate GWs in the
frequency range of $0.1$--$1 \, {\rm Hz}$. Recently, a creative concept for
a lunar-based laser-interferometer GW detector, GW Lunar Observatory for
Cosmology (GLOC)~\cite{Jani:2020gnz}, is proposed with a greater
sensitivity. For the first time, we will study in detail the prospects of
testing dipole radiation with those {\it ground-based} decihertz GW
detectors.

Multi-band observations from space-borne detectors and ground-based
laser-interferometer detectors have enormous abilities to test the dipole
radiation~\cite{Barausse:2016eii, Gnocchi:2019jzp, Carson:2019fxr,
Liu:2020nwz}. However, actual functioning schedules of these detectors will
introduce big uncertainties in performing multi-band tests, and to explore
as many possibilities as possible, in this work we consider ground-based
atom-interferometer GW detectors and ground-based laser-interferometer
detectors to test the dipole radiation. In principle, these
atom-interferometer GW observatories record the GW signals earlier than
AdvLIGO, CE, and ET, due to their lower operating frequencies. They can be
made use of to bound the dipole radiation tighter. Dipole radiation tests
can come from BBH systems or binary neutron star (BNS)
systems~\cite{Barausse:2016eii}. Both scenarios have unique roots in
specific theories. We focus on BNS systems in our study. On one hand, in
some classes of alternative theories of gravity, dipole GW emission from
BBHs is usually absent because of the no-hair
theorem~\cite{TheLIGOScientific:2016src, Berti:2015itd}. On the other hand,
as shown by \citet{Damour:1993hw}, some strong-field behaviors related to
dipole radiation only happen in neutron stars, and they cannot be studied
with BBHs. For some theories, probing dipole radiation from BNSs could
reveal unique properties of gravitation that BBHs hardly provide. This kind
of study is complementary to studies that probe the dipole radiation from
BBH systems in some other classes of gravity theories where BHs can be
scalarized.

In our study, we use one of the modest ground-based atom-interferometer
detectors, ZAIGA, as an example. As we will see, limited by the influence
of Newtonian noise, ZAIGA alone hardly detects GWs from distant BNS
sources. In addition, we consider an imaginary upgrade, what we call
``Z+'', whose strain sensitivity is improved by an order of magnitude
compared to that of ZAIGA (see e.g. Ref.~\cite{Chaibi:2016dze}). In total,
we consider AdvLIGO and its updates, CE, ET, ZAIGA, and Z+ in our study.
For statistical significance, we simulate ensembles of BNSs according to
the population properties of BNSs~\cite{LIGOScientific:2018mvr}. We perform
the parameter estimation to obtain the bounds on dipole radiation from
these systems for different detectors. As discussed earlier, due to the
lower accessible frequency of ZAIGA/Z+, the effect of $-1$\,PN correction
can be constrained rather well. The joint bounds show that, ZAIGA/Z+ and
other ground-based laser-interferometer detectors could complement each
other in testing the gravitational dipole radiation for BNS systems.
Especially, the constraints on the dipole radiation from AdvLIGO and CE can
be tighter with the help of ZAIGA/Z+. We consider this as an extra science
motivation for atom-interferometer detectors~\cite{Zhan:2019quq}.

The organization of the paper is as follows. In~\cref{sec:detectors}, we
briefly introduce GW observatories that are used in our study.
\Cref{sec:dipole} reviews the modified GW waveform with additional dipole
radiation in the parameterized post-Einsteinian (ppE) framework.
In~\cref{sec:method}, we consider the population property of BNSs and
simulate the possible BNS systems to be detected by different GW
observatories. Then, the distributions of bounds on the dipole radiation
parameter are obtained from the ensemble of BNSs. Finally, our main results
and discussions are given in~\cref{sec:res,sec:dis}, respectively.
Throughout the paper, we use the geometrized unit where $G = c = 1$.

\begin{table*}
    \caption{Relevant parameters for eight ground-based GW detectors
    in our study. Note that, the strain sensitivity $\sqrt{S_n}$
    of the imaginary upgrade of ZAIGA, Z+, is improved by an order of
    magnitude compared to that of ZAIGA; therefore, we have $S_{n, \, {\rm
    Z+}}=S_{n, \, {\rm ZAIGA}}/100$. \label{tab:paras}}
        \centering
        \def\arraystretch{1.3}
        \begin{tabularx}{\textwidth}{XXXXXXp{3cm}X}
          \hline\hline
       Detector & $f_{\rm min}\,(\rm Hz)$ & $f_{\rm max}\,(\rm Hz)$ & Shape & $f_\alpha$ & $n_\alpha$ & $S_{n,\alpha}$ & Schedule \\ \hline
        AdvLIGO & 5     & 5000 & Right-angled & 1            & 2 & LIGO document\footnote{\url{https://dcc.ligo.org/LIGO-T1800044/public}}  & O4~\cite{Abbott:2020qfu} \\
        A+LIGO  & 5     & 5000 & Right-angled & 1            & 2 & LIGO document\footnote{\url{https://dcc.ligo.org/LIGO-T1800042/public}}  & By 2025~\cite{Abbott:2020qfu}\\
        Voyager & 5     & 10000& Right-angled & 1            & 2 & LIGO document\footnote{\url{https://dcc.ligo.org/LIGO-T1500293/public}}  & Late 2020s~\cite{Adhikari:2019zpy}\\
       CE-1     & 3     & 10000& Right-angled & 1            & 2 & Refs.~\cite{Reitze:2019iox,Reitze:2019dyk}  & 2030s~\cite{Reitze:2019iox}\\
       CE-2     & 3     & 10000& Right-angled & 1            & 2 & Refs.~\cite{Reitze:2019iox,Reitze:2019dyk}  & 2040s~\cite{Reitze:2019iox}\\  
       ET-D     & 1     & 10000& Triangle     & $\sqrt{3}/2$ & 3 & Ref.~\cite{Hild:2010id}     & Mid 2030s~\cite{Maggiore:2019uih}\\
       ZAIGA    & 0.1   & 10   & Triangle     & $\sqrt{3}/2$ & 2 & Ref.~\cite{Zhan:2019quq}    & -- \\
        Z+      & 0.1   & 10   & Triangle     & $\sqrt{3}/2$ & 2 & $S_{n, \, {\rm ZAIGA}}/100$ & -- \\
       \hline
        \end{tabularx}
\end{table*}

\section{Detectors and signals}
\label{sec:detectors}

\begin{figure}
    \centering
    \includegraphics[width=8.0cm]{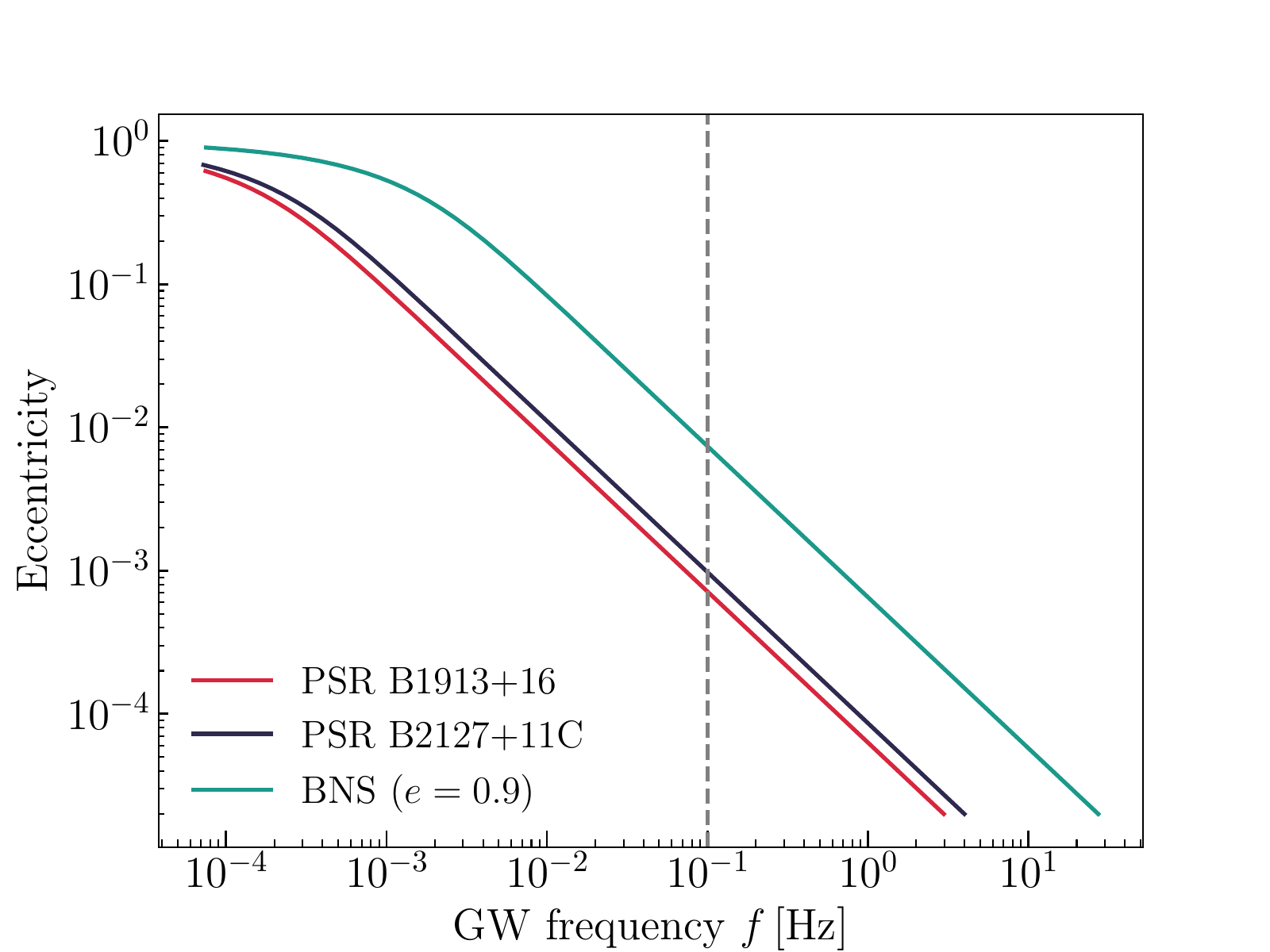}
    \caption{The evolution of eccentricity with GW frequency for three BNS systems~\cite{Kowalska:2010qg}.}
    \label{fig:ecc_f}
\end{figure}

\begin{figure*}
    \centering
    \includegraphics[width=12cm]{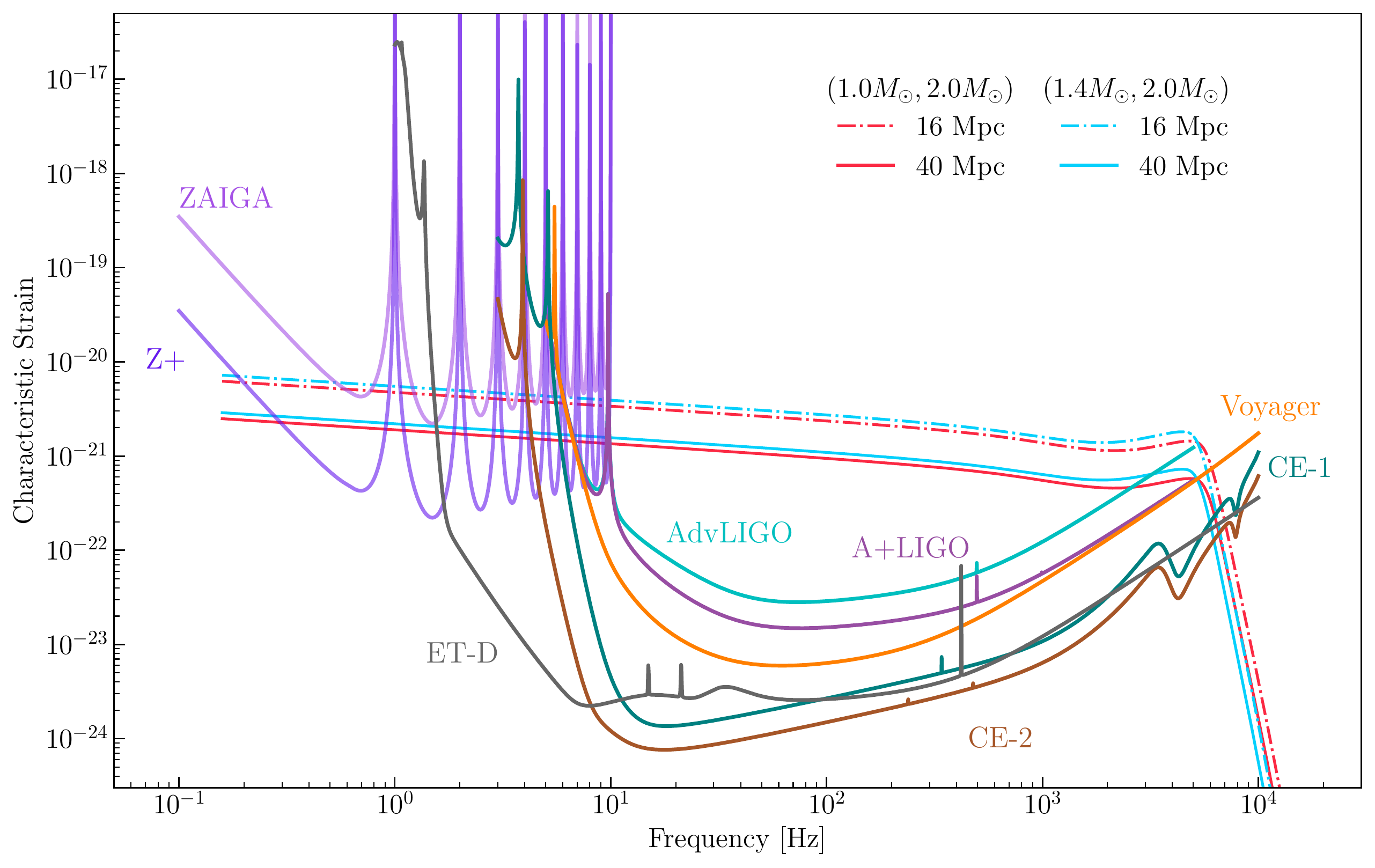}
    \caption{Noise curves for different detectors and characteristic
    strains for some example BNS systems. Note that we have taken account
    of the form factor and the equivalent number of detectors using
    $S_{n}^{\rm eff}$ in Eq.~(\ref{eq:Sneff}). Thus the SNR is related to
    the area between signal and noise curves~\cite{Moore:2014lga}. The GW
    waveform used for BNSs is the non-precessing BBH IMRPhenomD model as
    the high-frequency part has a minor effect in our dipole-radiation
    study. \label{fig:char_strain}}
\end{figure*}

GW observatories with a lower accessible frequency are more sensitive to
the $-1$\,PN dipole radiation. For the current GW detectors, AdvLIGO, AdV,
and KAGRA, their low frequencies are $\gtrsim 10 \, {\rm
Hz}$~\cite{Abbott:2020qfu}. Over the coming decades, they will be upgraded
to be more sensitive at low frequency. By 2025, the LIGO detectors will be
upgraded to A+LIGO~\cite{Abbott:2020qfu}. In the late 2020s, they will be
upgraded further to Voyager~\cite{Adhikari:2019zpy}. In the future, the
next-generation ground-based GW detectors, CE and ET, will further improve
their sensitivities. They are roughly an order of magnitude more sensitive
than AdvLIGO. CE will be realized in two stages: CE-1 in 2030s and CE-2 in
2040s~\cite{Reitze:2019iox,Reitze:2019dyk}. ET can extend the accessible
frequency down to $\sim 1 \, {\rm Hz}$. The design sensitivity of ET has
been updated a few times, and the latest configuration is known as
``ET-D''~\cite{Hild:2010id}.

For the atom-interferometer detector ZAIGA, its lower accessible frequency
is about $0.1 \, {\rm Hz}$. ZAIGA has two designs, the conservative ZAIGA
and the optimal ZAIGA that uses the same Newtonian noise subtraction as
ET~\cite{Zhan:2019quq}. In our work, we use the optimal one. In addition,
we notice that \citet{Chaibi:2016dze} proposed a method of exploiting the
Newtonian noises' correlation with an array of long baseline
atom-interferometer gradiometers. It is possible to achieve a {\it tenfold}
or more Newtonian noise rejection. Therefore, we further assume an
imaginary detector, that we call Z+, whose strain sensitivity is improved
by ten times comparing to ZAIGA. We stress that it is an optimal assumption
for now and we use it only for illustrative purposes.

In a summary, we use the following sets of ground-based detectors to
constrain dipole radiation: (i) laser-interferometer GW detectors, AdvLIGO
at its design sensitivity, A+LIGO, Voyager, CE-1, CE-2, and ET-D, and (ii)
ground-based atom-interferometer GW detectors, ZAIGA and Z+. These
detectors consist of two different types of shapes, the right-angled shape,
and the triangle shape. For a triangle detector $\alpha$, there is an
additional {\it effective} form factor $f_{\alpha} = \sqrt{3}/2$ compared
to an equivalent right-angled detector. Besides, a triangle detector is
usually equivalent to two independent detectors.\footnote{Note that a
triangle ET is equivalent to three independent right-angled detectors
because of its specific design~\cite{Hild:2010id,Freise:2008dk}.} For a
uniform treatment in the calculation, we absorb the form factor,
$f_{\alpha}$, as well as its equivalent number of detectors, $n_\alpha$,
into an {\it effective} power sensitivity density, $S_{n, \alpha}^{\rm
eff}$.\footnote{For a right-angled detector $\alpha$, we simply have
$f_{\alpha} = 1$ and $n_{\alpha}$ equals to their actual number of
detectors.} Thus, for a detector $\alpha$, we have
\begin{equation}
    S_{n, \alpha}^{\rm eff} = \frac{1}{f_\alpha^2 n_\alpha} S_{n,
    \alpha}\,. \label{eq:Sneff}
\end{equation}
For different detectors we use, we list their relevant configurations,
frequency range $\left[f_{\rm min}, \, f_{\rm max}\right]$, form factor
$f_\alpha$, equivalent detector number $n_\alpha$, and the scheduled epochs
when those detectors are to be operational in~\cref{tab:paras}.

For the BNS waveform, we use the inspiral-merger-ringdown phenomenological
frequency-domain GW waveform, the so-called IMRPhenomD
waveform~\cite{Husa:2015iqa, Khan:2015jqa}. A complete GW response for a
detector is $\widetilde{h}(f) = F^{+} \widetilde{h}_{+}(f) + F^{\times}
\widetilde{h}_{\times}(f)$, where $F^{+,\times}$ are the detector's
``+/$\times$'' component pattern functions related to the sky location
$(\theta, \phi)$, and the polarization angle $\psi$ of the GWs. The strain
$\widetilde{h}_{+,\times}$ can be obtained from the IMRPhenomD model. The
difference between BNSs and BBHs in GW waveform mainly comes from the
merger and ringdown parts induced by BNSs' matter
effects~\cite{Abbott:2018lct}. It is highly related to the BNSs' equation
of state. However, for BNSs the frequencies corresponding to the merger and
ringdown are too high to contribute to the dipole radiation test in a
noticeable way. Therefore, although the IMRPhenomD waveform is built for
BBHs, we consider it sufficiently accurate to be used to estimate the
dipole radiation parameter for BNSs.

The IMRPhenomD waveform assumes quasi-circular
inspirals. According to the formation mechanism of BNSs~\cite{Tauris:2017omb},
there could be a significant fraction of eccentric BNS binaries. If the orbital
eccentricity of a BNS is non-negligible, the GW amplitudes will split into
harmonics with comparable strength, and thus the IMRPhenomD waveform will not be
suitable~\cite{Peters:1963ux}. To explore the effects from eccentricity, we illustrate the evolution of
eccentricity with GW frequency for three BNS systems in~\cref{fig:ecc_f}. Two of them, PSRs B1913+16~\cite{Weisberg:2016jye}
and B2127+11C~\cite{Jacoby:2006dy}, have been long-time monitored by radio
telescopes. We can take PSR B1913+16 as an
example. Although the eccentricity of PSR B1913+16 is currently $e=0.617$ (the
orbital period is $7.75$ hours), it will have an eccentricity $e< 10^{-3}$
when its GW frequency $f = 0.1 \, {\rm Hz}$.  Furthermore, even if we consider a BNS
system ``BNS ($e=0.9$)'', a hypothetical BNS with an orbital period $P_{b} = 8.0
\, {\rm hours}$ and an extreme eccentricity $e=0.9$, the
eccentricity at $f \sim 0.1 \, {\rm Hz}$ is tiny enough to be neglected.  In
addition, we calculate the number of GW inspiral cycles contributed by
eccentricity, $\Delta N_{\rm GW}^{\rm ecc} = -7 \times 10^{6} \,
e_0^2$ where $e_0$ is the eccentricity at $f=0.1\,$Hz~\cite{Moore:2016qxz}. We find $\Delta N_{\rm GW}^{\rm ecc}$ is
quite small
 relative to the Newtonian contribution $N^{\rm Newt}_{\rm GW} \sim 1 \times
10^{7}$. Therefore, the orbital eccentricity can be assumed to be zero in our
consideration.

The effective detector noise amplitude is $h_n = \sqrt{f S_{n}^{\rm eff}}$
and the characteristic strain amplitudes of GWs are $h_c = 2f \left|
\widetilde{h} \right|$~\cite{Moore:2014lga}. In~\cref{fig:char_strain}, we
illustrate the $h_n$ for different ground-based detectors. In addition, we
plot the $(\theta, \phi, \psi)$-averaged and face-on ($\iota=0$)
characteristic strain $h_c$ of GWs from typical BNS systems.

The signal-to-noise ratio (SNR) $\rho$ for a detector can be expressed as
$\rho = \left(\widetilde{h}(f) \right| \left. \widetilde{h}(f) \right)^{1/2}$,
where the noise-weighted inner product is defined via,
\begin{equation}
    \label{eqn:matched-filtering}
    \left(\widetilde{A}(f) \right| \left. \widetilde{B}(f) \right)
    \equiv 2 \int_{f_{\rm min}}^{f_{\rm max}} \frac{\widetilde{A}^{\star}(f) \widetilde{B}(f) + \widetilde{A}(f) \widetilde{B}^{\star}(f)}{S_{n}^{\rm eff}} \rmd f \,.
\end{equation}
In~\cref{fig:char_strain}, the area between $h_n$ and $h_c$ curves is
proportional to the SNR~\cite{Moore:2014lga}. As we can see, a $(1.4
M_{\odot},\,2.0 M_{\odot})$ BNS at a luminosity distance $\DL=40 \, {\rm
Mpc}$ (at the same distance as the first detected BNS inspiral GW170817)
can hardly be detected by ZAIGA alone as it has a too small SNR. Only very
close BNSs are detectable by ZAIGA-like detectors.

\begin{figure}
    \centering
    \includegraphics[width=7.5cm]{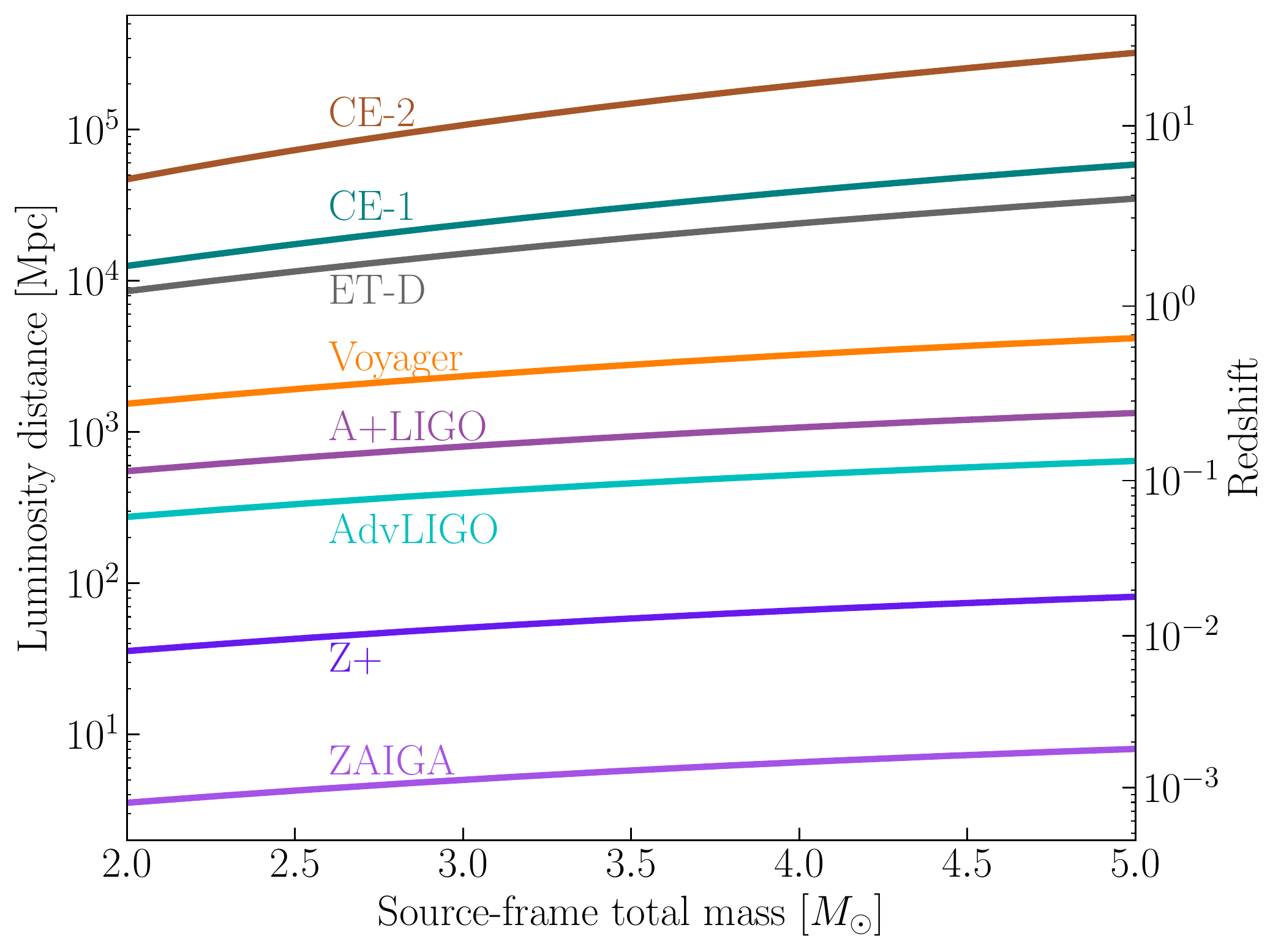}
    \caption{The luminosity distance ({\it left}) of equal-mass face-on
    BNSs and the corresponding redshift ({\it right}) with a SNR $\rho=8$
    for ground-based GW detectors. We have averaged the sky location and GW
    polarization angles. \label{fig:horizon}}
  \end{figure}
\begin{figure}
    \centering
    \includegraphics[width=6cm]{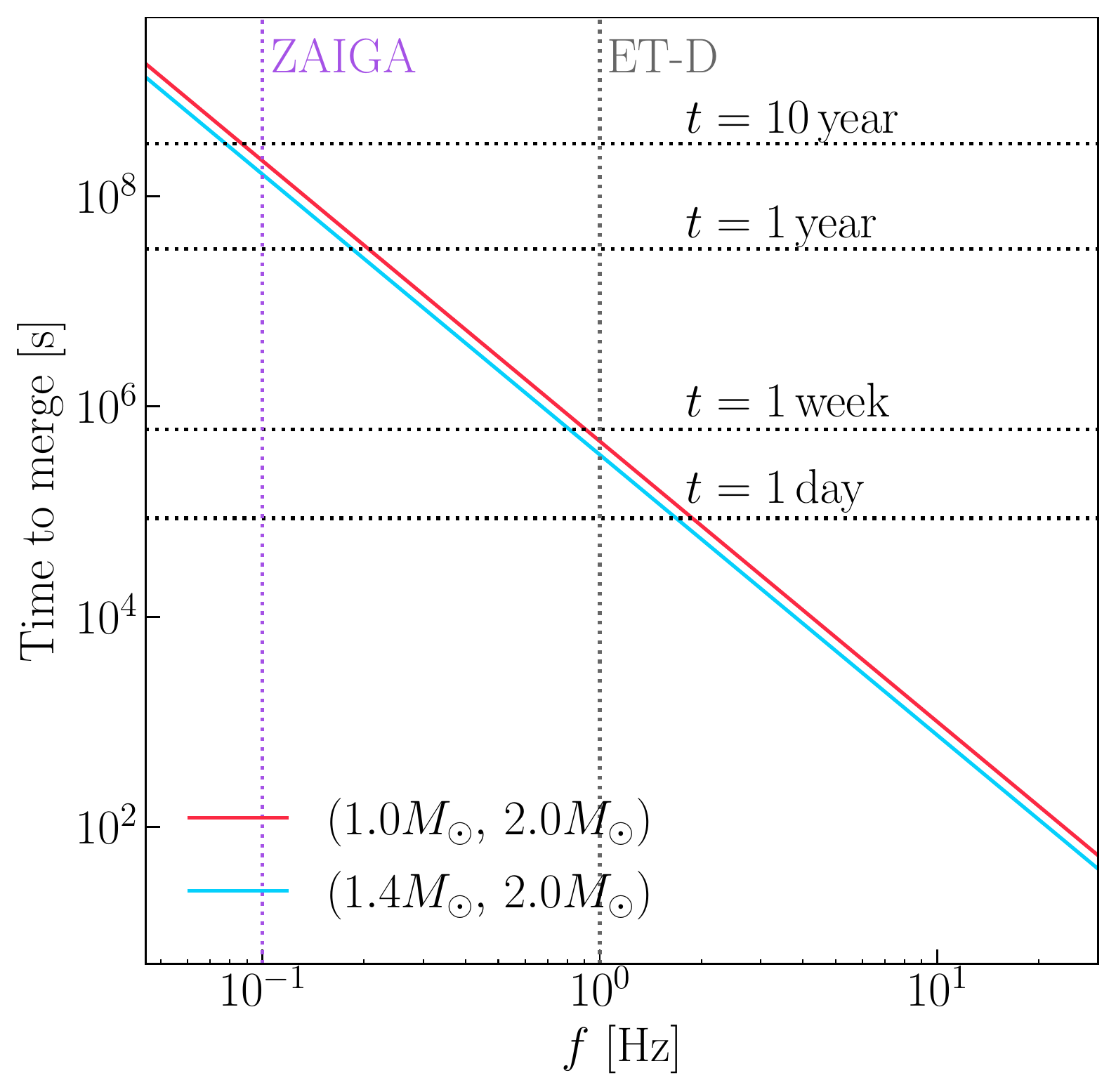}
    \caption{Time to coalescence for two example BNS sources. The vertical
    dotted lines correspond to the lower frequency of ZAIGA/Z+ and ET-D.
    \label{fig:time2merge}}
\end{figure}

We illustrate the range of luminosity distances for equal-mass face-on BNSs
with different detectors assuming a threshold SNR $\rho=8$
in~\cref{fig:horizon}. In the figure, the sky location of binaries and GW
polarization angle have been averaged. In addition, we adopt a cosmology
with the matter density $\Omega_{\rm M} \simeq 0.315$, the dark energy
density $\Omega_\Lambda = 1- \Omega_{\rm M} \simeq 0.685$, and the Hubble
constant at today $H_0 \simeq 67.4 \, {\rm km \, s^{-1} \,
Mpc^{-1}}$~\cite{Aghanim:2018eyx} to convert between the source redshift
$z$ and its luminosity distance $\DL$, where the luminosity distance as a
function of redshift $z$ is given by $\DL(z) = (1+z) \int_{0}^{z} \rmd
z^{\prime} / {H(z^{\prime})} $, with $H(z) \equiv H_{0}
\sqrt{\Omega_{\mathrm{M}}(1+z)^{3} + \Omega_{\Lambda}}$. As we can see
in~\cref{fig:horizon}, ZAIGA can hardly detect GWs alone from BNSs whose
luminosity distances are larger than $10 \, {\rm Mpc}$. For Z+, its reach
to BNSs is less than $100 \, {\rm Mpc}$, but it is already within the
astrophysically interesting range.\footnote{Note that the first detected
BNS inspiral, GW170817, is certainly within Z+'s reach.} Among the
detectors we use, CE-2 can detect the furthest BNSs, to a distance as far
as $\sim 100 \, {\rm Gpc}$.

The frequency of GW, $f$, evolves at the Newtonian order via ${\dot{f}} /
{f} ={96} {\left(\pi \calM f \right)^{8 / 3}}/ 5{\calM}$, for a binary
system, where $\calM=(m_1 m_2)^{3/5}/(m_1 +m_2)^{1/5}$ is the redshifted
chirp mass with redshifted component masses $m_{1}$ and $m_2$. Integrating
it from an initial frequency $f_{\rm in}$, one obtains the time to
coalescence, $\Delta\tau(f_{\rm in}) \approx {5} \calM \left(\pi \calM
f_{\rm in} \right)^{-8/3} / {256} $. We show the time to coalescence at
$z=0$ for typical BNSs in~\cref{fig:time2merge}. As we can see, for a
$(1.0\, M_{\odot}, 2.0\,M_{\odot})$ binary system, the GW signals stay for
a very long time. For ET, it can last a week before its coalescence. For
ZAIGA/Z+, we can record the GW signal for more than 1 year. This shows
exactly the point that a lower accessible frequency can be advantageous to
test the negative PN effects in BNS systems, such as the dipole radiation.

\section{Dipole radiation in BNS systems}
\label{sec:dipole}

One of the earliest motivations to test dipole radiation comes from the
Jordan-Fierz-Brans-Dicke (JFBD) theory~\cite{Fujii:2003pa}. This theory
adds a scalar degree of freedom in addition to the metric field in GR.
Famous extensions include the Damour-Esposito-Far\`ese
theory~\cite{Damour:1993hw, Damour:1996ke}, which predicts strong-field
scalarization for NSs and has been tested by a couple of binary
pulsars~\cite{Freire:2012mg, Wex:2014nva, Shao:2017gwu, Zhao:2019suc, Guo:2021leu}. The
constraints on dipole radiation from binary pulsars are rather tight now in
the quasi-stationary strong-field regime~\cite{Wex:2014nva}. For the
highly-dynamical strong-field regime, compact binary coalescences become
one of the most powerful testbeds. For example, certain kind of theories
(e.g. with a scalar Gauss-Bonnet coupling term~\cite{Berti:2015itd})
predict scalar hairs for BHs, and the dipole radiation in these theories
can be constrained with BBH systems~\cite{Liu:2020nwz, Barausse:2016eii,
Barausse:2017gip}. The specifics in the underlying gravity theory determine
whether NSs or BHs get scalarized. In this paper, with the
Damour-Esposito-Far\`ese gravity~\cite{Damour:1993hw, Damour:1996ke} kept
in mind as a prototype theory, we use GWs from BNS systems to test the
dipole radiation. But the analysis can be extended to other scenarios as
well when needed.

The extra scalar field not only causes an additional gravitational binding
energy shift for the orbit but also enhances the energy flux via extra
dipole radiations. Generically, the modified GW flux can be parameterized
as~\cite{Barausse:2016eii}
\begin{equation}\label{eq:B}
    \dot{E}^{\rm GW} = \dot{E}^{\rm GR} \left[ 1 + B \left( \frac{m}{r} \right)^{-1} \right] \,,
\end{equation}
where $\dot{E}^{\rm GR}$ is the GW flux derived from GR's quadrupolar
radiation, $m$ and $r$ are respectively the total mass and the orbital
separation of the binary, and the theory-dependent parameter $B$ describes
the strength of the additional dipole radiation. Compared to the
quadrupolar radiation in GR, dipole radiation is enhanced by a factor of
$\sim ( m/r )^{-1} \sim (v/c)^{-2} $, where $v$ is the characteristic
relative speed in the binary. As said before, the dipole emission is a
$-1$\,PN effect, and dominates relatively at a larger separation $r$, or at
a lower GW frequency. For JFBD-like scalar-tensor theories, $B$ equals to
$(5/96) |\Delta \alpha|^2$, where $\Delta \alpha$ describes the difference
between the effective scalar couplings of two objects in the
binary~\cite{Damour:1993hw,Tahura:2018zuq}.

To incorporate the effects from dipole radiation into GW waveform, we use
the ppE framework~\cite{Yunes:2009ke, Tahura:2018zuq}. This framework
generically parameterizes GW waveform's deviations from the GR prediction.
The GW waveform for a compact binary inspiral in the ppE framework is given
by,
\begin{equation}
    \label{eqn:ppE}
    \htilde(f) =  \htilde^{\rm GR}(f) \cdot \left( 1 + \alpha u^{a} \right) e^{\ii \beta u^{b}} \,,
\end{equation}
where $u \equiv \pi \calM f$, and $\htilde^{\rm GR}(f) $ is the waveform in
GR. The amplitude and phase of GW waveform are corrected by the ppE
parameters $(a, \alpha)$ and $(b, \beta)$
respectively~\cite{Yunes:2009ke,Tahura:2018zuq}. In particular,
Eq.~(\ref{eqn:ppE}) reduces to the waveform in GR with $\alpha=\beta=0$.
For theories with dipole radiation, the general modification to the
inspiral phase is given by $b = -7/3$, and $\beta$ is connected to the
dipole radiation parameter $B$ in Eq.~(\ref{eq:B}) via $-(3/224) \eta^{2/5}
B$. Note that, the non-GR corrections for the GW amplitude are less
important in general, as the matched filtering is more sensitive to the
phase evolution. Having neglected the matter effects at high frequency in
the merger-ringdown stages, we use the IMRPhenomD BBH waveform to represent
$\htilde^{\rm GR}(f)$ in~\cref{eqn:ppE}. Also, when we consider the ppE
modified waveform, it is sufficient to only include the inspiral phase
corrections in our current problem.

\section{Testing dipole radiation with BNS populations}
\label{sec:method}

In our test, the dipole radiation is parameterized by the parameter $B$ in
Eq.~(\ref{eq:B}). Before
investigating the bounds on the parameter $B$ with the Fisher information matrix
(FIM) method, we can use the number of GW inspiral cycles, $N_{\rm GW}$, for a
rough estimate of bounds on $B$~\cite{Will:1994fb}. Here, we take a typical BNS
system as an example. For a $(1.4 M_\odot, 1.4 M_\odot)$ BNS system
detected by ZAIGA/Z+, the number of GW inspiral cycles contributed by dipole
radiation is $ \Delta N_{\rm GW}^{\rm dipole} \sim - 2 \times 10^{10} \, B$.
When we set $ \left| \Delta N_{\rm GW}^{\rm dipole} \right | \lesssim 1$, we can obtain
a rough estimation, $B \lesssim 5 \times 10^{-11}$. In addition, we use another
method from Ref.~\cite{Barausse:2016eii}, using $\left|(\Delta t)_{\rm GR}-(\Delta t)_{\rm non-GR}\right| $ to
roughly bound $B$. The result is consistent with the
constraint from $ \left| \Delta N_{\rm GW}^{\rm dipole} \right |< 1$. Note
that, these simple methods above only involve parameter $B$, ignoring 
correlations between $B$ and other parameters. For more complete analyses, we will use
the FIM for a more reasonable estimation.\footnote{Comparing
$\Delta N_{\rm GW}^{\rm dipole}$ and $\Delta N_{\rm GW}^{\rm ecc}$, it might be possible
to discuss whether one can neglect the eccentricity in ppE waveforms
though this is a simple analysis. For a better analysis, one might
apply the same analysis in Ref.~\cite{Favata:2013rwa} to eccentric ppE waveforms based on
Ref.~\cite{Loutrel:2014vja}.}

The FIM is constructed as usual from the frequency-domain waveform
$\htilde(f)$,
\begin{equation}
    \label{eqn:fisher}
    \Gamma_{i j} \equiv \left(\frac{\partial \htilde(f)}{\partial {\xi}_i }  \right| 
            \left. \frac{\partial \htilde(f)}{\partial {\xi}_j } \right) \,.
\end{equation}
Here, the parameter $\xi_i \in \left\{ \ln \calA, \ln \calM, \ln \eta,
\chi_1, \chi_2, \phi_c, t_c, B \right\}$ where $\calA$ is the GW amplitude, $\eta=m_1
m_2/\left(m_1+m_2\right)^2$ is the symmetric mass ratio of the binary, $\phi_c$
and $t_c$ are the phase and time at coalescence respectively, and
$\chi_{1,2}$ is the dimensionless aligned spin components of two NSs. For a
parameter ${\xi}_i$, as per the Cram\'er-Rao inequality, a lower bound on
its standard deviation expected from an experiment can be placed with
$\sigma\left(\xi_i\right) \geq \sqrt{ \left(\Gamma^{-1} \right)_{ii}
}$~\cite{Cramer:1946,Rao:1992,Vallisneri:2007ev}. We refer the readers to
\cref{sec:app:FIM} for a brief explanation of the FIM method.

To have a statistically sound estimate, we consider the population property
of BNSs and simulate the possible BNS systems across the cosmos. First, we
denote the BNS merger rate per unit proper time per unit co-moving volume
at $z=0$ by $\mathcal{R}_0$. It is convenient to have BNS merger rate at
redshift $z$ using,
\begin{equation}
    \mathcal{R}(z) = \mathcal{R}_0 \, r(z) \,.
\end{equation}
Here, the merger rate's time evolution is encapsulated in $r(z)$. We use
the same piecewise linear fitting to $r(z)$ as that in
Refs.~\cite{Cutler:2005qq,Schneider:2000sg}. In that way, $r(z)$ can be
expressed using the following piecewise function,
\begin{equation}
    r(z)=
    \begin{cases}
        1+2 z \,, & \quad z < 1 \,, \\
        \frac{3}{4}(5-z) \,, & \quad 1 \leq z \leq 5 \,, \\
        0 \,, & \quad z > 5 \,.
    \end{cases}
\end{equation}
This is a rough approximation to the realistic BNS population, but suffices
for our study here.

Over a fixed observing duration $\Delta T$, the total number of BNS merger
events, $N_{\rm event}$, detected by a single detector is,
\begin{equation}
    \label{eqn:event}
    N_{\rm event}= \Delta T \times \int_{0}^{z_{\rm th}} \frac{4 \pi \, D_{\rm COM}^{2}(z) \, \mathcal{R}(z)}{(1+z) \, H(z)}\rmd z \,,
\end{equation}
where $z_{\rm th}$ is the redshift corresponding to response distance for a
detector~\cite{Chen:2017wpg} and $D_{\rm COM}(z) = \DL(z)/(1+z)$ is the
co-moving distance to the BNS system. Here, for each detector we set
$\Delta T = 1 \, {\rm year}$, as a conservative operational duration, to
derive the total numbers of BNS merger events.\footnote{Correspondingly, for
ZAIGA/Z+, considering the operational duration, we actually use the frequency
one year before coalescence, $f_{\rm min} = f_{\rm 1yr}$, in
\cref{eqn:matched-filtering}. For a $(1.4 M_{\odot}, 1.4 M_{\odot})$ BNS system at $z=0$,
$f_{\rm 1yr} \sim 0.2 \, {\rm Hz}$.}

In reality, many BNSs will be detected by the aforementioned GW detectors
except ZAIGA. Some of these BNSs will be of low SNRs, while the others are
rather loud. Depending on the specific systems' physical parameters, these
BNSs will provide different limits on the dipole radiation parameter $B$.
The ultimate constraint is to be dominated by the best limit, or a
combination of a few best ones. Therefore, in this paper, we focus on one or
a few best bounds from the BNS population. We randomly generate BNS
populations from an underlying population model and calculate the optimal
uncertainty of $B$ in each population. Such a treatment differs from some
earlier studies but is close to reality when a test of dipole radiation is
to be carried out.

For some detectors (e.g., Voyager, CE, and ET), their detection ranges,
denoted by the redshift $z_{\rm th}$, are very large. The computational
cost for the whole population is expensive. As we have explained, the
optimal constraints are to be obtained from the closer binary systems to
us. Therefore, instead of using the threshold redshift $z_{\rm th}$ in
Eq.~(\ref{eqn:event}), we use $z = 0.5$ for Voyager, CE, and ET detectors.
It is enough for calculating the optimal bounds on $B$ in our study. For
ZAIGA, its $z_{\rm th}$ is too small to detect a realistic BNS system
unless we are very lucky. Therefore, we set an over-optimal $z_{\rm th} = 0.02$ for ZAIGA, which is
the same value as that from Z+. Note that, the sensitivity of ZAIGA is
still worse than that of Z+. This approach only ensures that ZAIGA will have some (low-SNR) events to be analyzed; see more details in the following text about the usage of ZAIGA with other ground-based laser interferometers. The threshold redshift $z_{\rm th}$ that we use
in the simulation and the corresponding total numbers of BNS merger events
per year, $N_{\rm event}$, are given in~\cref{tab:population}.

\begin{table*}
    \caption{Relevant parameters for the simulation of BNS populations and the
    median bounds on the relevant parameters. For $N_{\rm event}$ we list the
    number of BNSs that have $\rho \gtrsim 8$. Note that, ``*'' stands for the
    bounds on parameters using an `improper' FIM, and the results are roughly
    suitable only for the order-of-magnitude estimates. \label{tab:population}}
    \centering
    \def\arraystretch{1.3} 
    \begin{tabularx}{0.82\textwidth}{p{1.5cm}p{1.2cm}p{1.5cm}p{2cm}p{2cm}p{2cm}p{2cm}p{2cm}}
        \hline\hline
        Detector & $z_{\rm th}$   & $N_{\rm event}$ & $\sigma_{\rm opt}(B)$
        (median) & $\sigma_{\rm com}(B)$ (median) & $\sigma_{\rm opt}(
        \mathcal{M})/\mathcal{M}$  (median) & $\sigma_{\rm opt}(
        \eta)/\eta$  (median) & $\sigma_{\rm opt}(t_c) \,{\rm [s]}$  (median)  \\ \hline
  
        ZAIGA   & $<0.02$         & $<4$         & $1.3 \times 10^{-8}$* & $9.8 \times 10^{-9}$* & $1.2 \times 10^{-5}$* & $4.0 \times 10^{-3}$* & $6.4 \times 10^{-1}$*  \\
  Z+      & $\lesssim 0.02$ & $\lesssim 4$ & $1.2 \times 10^{-9}$* & $9.4 \times 10^{-10}$* & $1.1 \times 10^{-6}$* & $3.9 \times 10^{-4}$* & $6.0 \times 10^{-2}$*  \\
  AdvLIGO & $\lesssim 0.2$  & $\sim 35$    & $7.2 \times 10^{-7}$ & $3.0 \times 10^{-7}$ & $5.8 \times 10^{-5}$ & $8.9 \times 10^{-3}$ & $7.1 \times 10^{-5}$  \\
  A+LIGO  & $\lesssim 0.3$  & $\sim 280$   & $4.8 \times 10^{-7}$ & $2.0 \times 10^{-7}$ & $3.5 \times 10^{-5}$ & $4.6 \times 10^{-3}$ & $3.4 \times 10^{-5}$  \\
  Voyager & $\sim0.5$       & $\sim 5600$  & $2.1 \times 10^{-7}$ & $8.9 \times 10^{-8}$ & $1.6 \times 10^{-5}$ & $2.0 \times 10^{-3}$ & $2.3 \times 10^{-5}$  \\
  CE-1    & $>0.5$          & $>47551$     & $2.8 \times 10^{-8}$ & $1.2 \times 10^{-8}$ & $2.2 \times 10^{-6}$ & $1.9 \times 10^{-4}$ & $5.1 \times 10^{-6}$  \\
  CE-2    & $>0.5$          & $>47551$     & $8.4 \times 10^{-9}$ & $3.5 \times 10^{-9}$ & $7.8 \times 10^{-7}$ & $1.0 \times 10^{-4}$ & $2.7 \times 10^{-6}$  \\
  ET-D    & $>0.5$          & $>47551$     & $2.1 \times 10^{-9}$ & $8.7 \times 10^{-10}$ & $3.3 \times 10^{-7}$ & $5.6 \times 10^{-5}$ & $4.2 \times 10^{-6}$ \\
        \hline
    \end{tabularx}
\end{table*}

In the simulation, we generate $N_{\rm event}$ possible BNS systems within
one-year time according to~\cref{eqn:event}. Meanwhile, the angle
parameters for the sky location, $\cos \theta$ and $\phi$, are generated
randomly from $[-1, 1]$ and $[0, 2 \pi)$, respectively. The GW polarization
angle $\psi$ is uniform in $[0, 2 \pi)$ and the inclination angle $\cos
\iota$ is generated uniformly in $[-1, 1]$. The source-frame masses of
binaries are obtained in Gaussian distribution $\mathcal{N}\left(1.35 \,
M_{\odot},\, \left(0.09 \, M_{\odot}\right)^{2}
\right)$~\cite{LIGOScientific:2018mvr, Ozel:2016oaf}. A ``low spin'' BNS
population, where the aligned spin components $\chi_{1,2}$ are uniform in
$[-0.05,0.05]$, is assumed in our computation. A recent merger rate of BNS,
$\mathcal{R}_0 = 1210 \, {\rm Gpc^{-3}\,year^{-1}}$, is used in the
simulation~\cite{LIGOScientific:2018mvr}.

According to~\cref{tab:population}, we can see that at least thousands of
BNS systems could be detected in AdvLIGO, A+LIGO, Voyage, CE and ET.
However, only the few ones that give the best constraints on dipole
radiation contribute to the final bounds on $B$ in our simulation, while
the majority of the remaining BNS samples hardly contribute. As we know,
for a given detector, the FIM is roughly proportional to the square of SNR
$\rho^2 \propto \calA^2$, according to~\cref{eqn:fisher}. In this way, we
expect the bounds on the dipole radiation parameter $\sigma(B) \propto
\rho^{-1} \propto \calA^{-1}$ in a rough manner. Therefore, to simplify the
computation further, we construct an {\it effective amplitude} in
proportional to the original $\calA$,
\begin{equation}
    \mathsf{A} = \left[\left(1+\cos^{2} \iota \right)^{2} F_{+}^{2}+4 \cos^{2} \iota \, F_{\times}^{2}\right]^{1 / 2} \frac{\calM^{5/6}}{\DL} \,,
\end{equation}
to reduce the total number of BNS mergers in the FIM calculation. We use
$\mathsf{A}$ to get the loudest BNS systems in advance. Then, instead of
using all possible BNS mergers, only the loudest BNS systems associated
with relatively larger $\mathsf{A}$'s are considered in our FIM
calculation. For a detector, we choose the total number of the loudest BNS
systems to be $N= \min \left( N_{\rm event},\, N_{\rm cut} \right)$. Here,
we set $N_{\rm cut}=30$.\footnote{For a negative PN effect (here, dipole
radiation effect), the combined
bounds are usually from a few tightest bounds. Our setting $N_{\rm cut}=30$ is
consistent with the results of Ref.~\cite{Perkins:2020tra}.} In other words, if $N_{\rm event} > N_{\rm cut}$,
we will only use the {\it top} $N_{\rm cut}$ BNS systems corresponding to
large $\mathsf{A}$'s. According to the formula for combination (see below),
it is enough to obtain the constraints from those BNS systems. We record
the tightest constraint, $\sigma_{\rm opt}(B)$. In addition, the combined
constraint from $N$ events, denoted as $\sigma_{\rm com}(B)$, is obtained
via,
\begin{equation}
    \frac{1}{\sigma_{\rm com}^2(B)} = \sum^{N}_{i=1}
    \frac{1}{\sigma_i^2(B)} \,.
\end{equation}

We repeat  simulations with random samples  $1000$ times for
each GW detector. At the end, we collect results of $\sigma(B)$.
The distributions of  1000 ``optimal'' bounds, $\sigma_{\rm opt}(B)$, and 1000 ``combined'' bounds, $\sigma_{\rm com}(B)$,
 can be obtained from the above simulations. Moreover,
 to investigate how laser interferometers and atom
 interferometers will complement each other in testing the dipole radiation in BNS
 systems, we take two scenarios as follows,
\begin{itemize}
  \item Scenario (I): investigating the bounds on the dipole radiation
  parameter from individual ground-based detectors;
  \item Scenario (II): investigating the joint bounds on $B$ by
  strategically combining atom interferometers, ZAIGA/Z+, with laser
  interferometers.
\end{itemize}
In different scenarios, we perform the parameter estimation with FIM and
obtain the desired bounds, which are presented in the next section.

\section{Projected constraints on dipole radiation}
\label{sec:res}

\begin{figure}
    \centering
    \includegraphics[width=8cm]{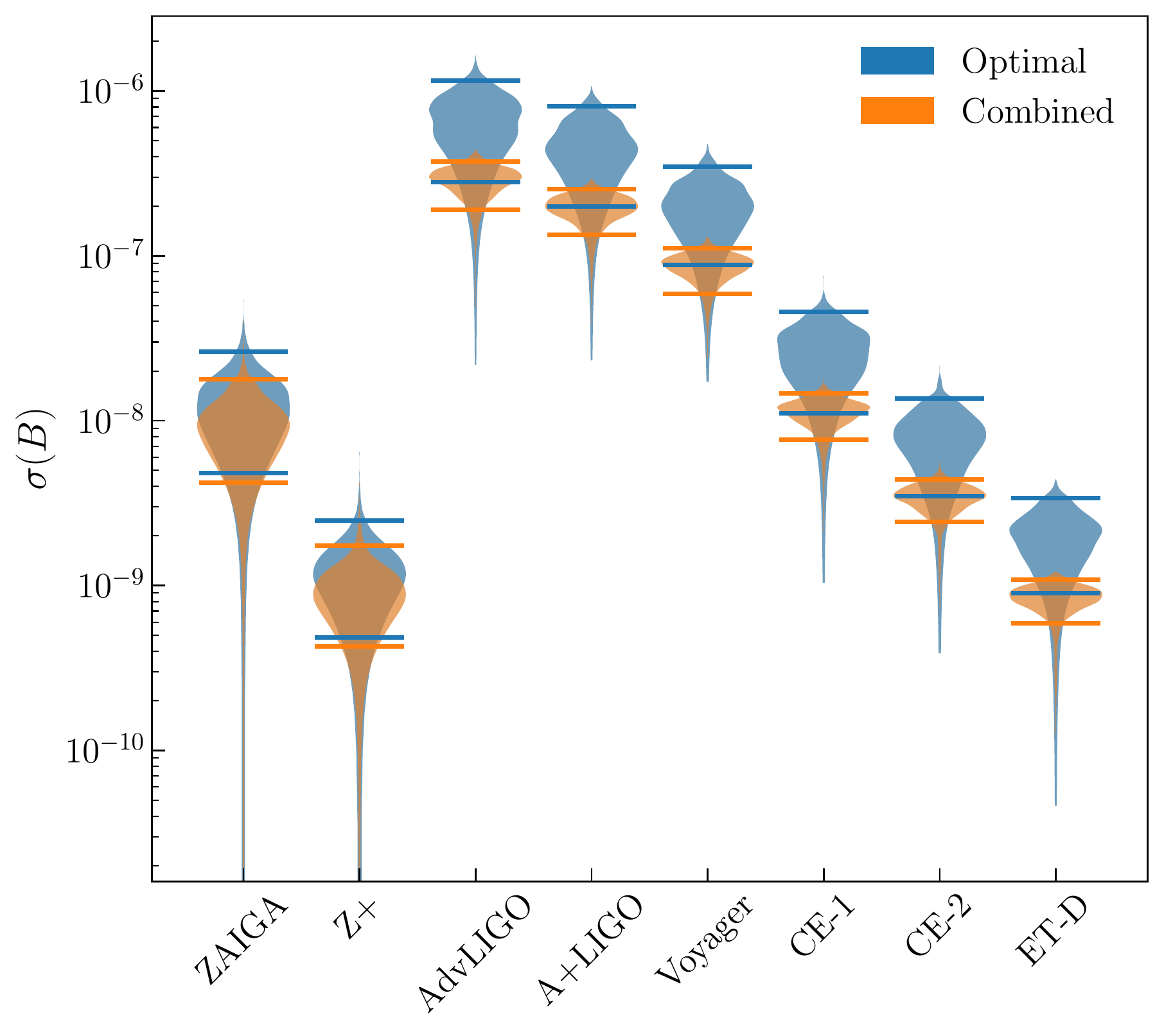}
    \caption{The distribution of $\sigma_{\rm opt}(B)$ (blue) and
    $\sigma_{\rm com}(B)$ (orange) from eight ground-based detectors in
    Scenario (I). The upper and lower bars show the values at 5\% and
    95\% confidence levels. \label{fig:individual_90CL}}
\end{figure}

\begin{figure*}
    \centering
    \hspace{-0.3cm}\includegraphics[width=12cm]{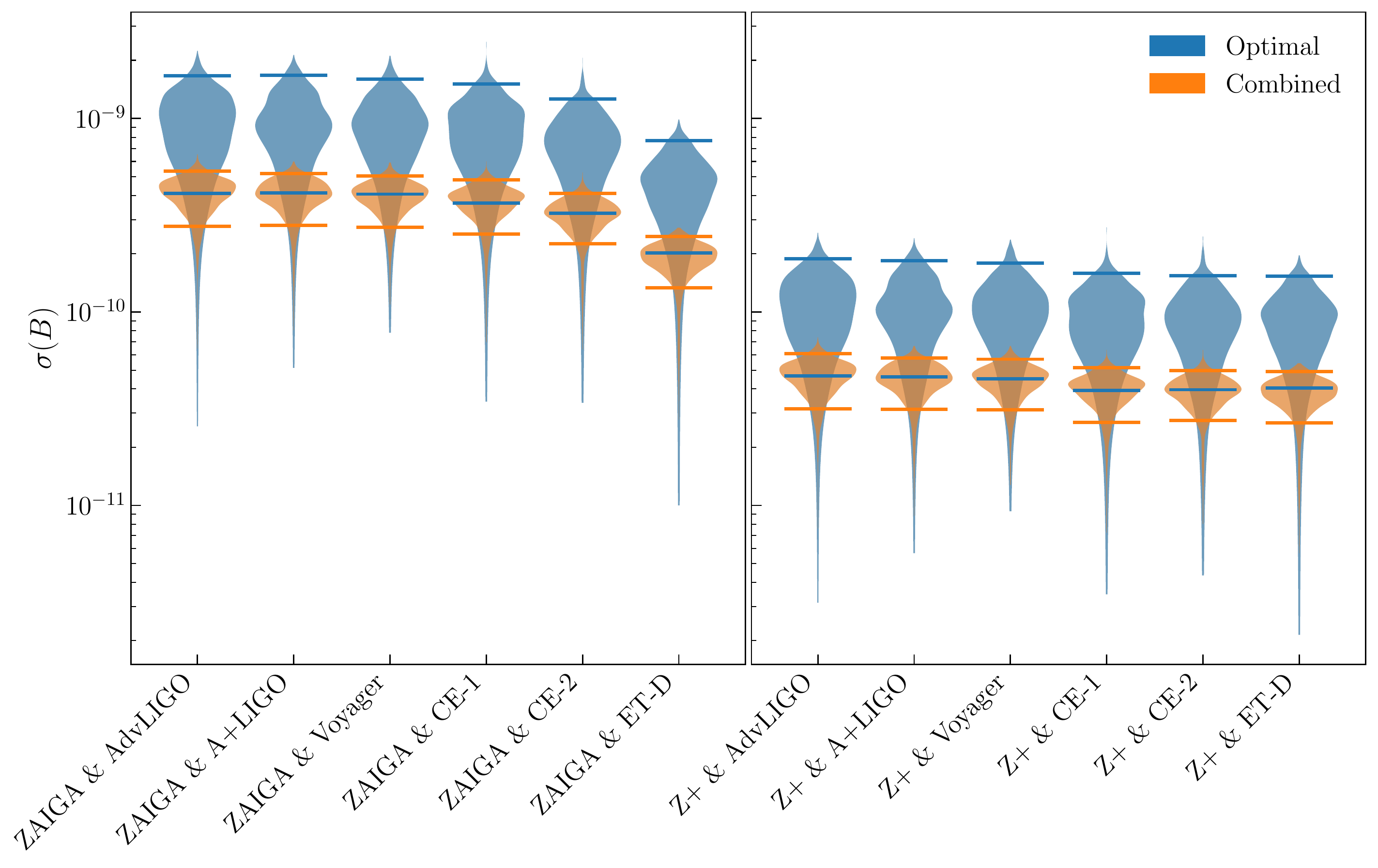}
  \caption{Same as~\cref{fig:individual_90CL}, but for Scenario (II), where
  the FIMs of laser interferometers are combined with that of atom
  interferometers ZAIGA (left) and Z+ (right). \label{fig:joint_90CL}}
\end{figure*}

From the setup of simulations in the last section, we obtain the
distributions of the parameter $\sigma(B)$, including $\sigma_{\rm opt}(B)$
and $\sigma_{\rm com}(B)$. They are illustrated
in~\cref{fig:individual_90CL,fig:joint_90CL}, and will be explained
further in the following. For each detector, the median values of
$\sigma(B)$ from the distributions are collected in~\cref{tab:population}
for references.

\begin{figure}
    \hspace{-0.3cm}
    \includegraphics[width=0.5\textwidth]{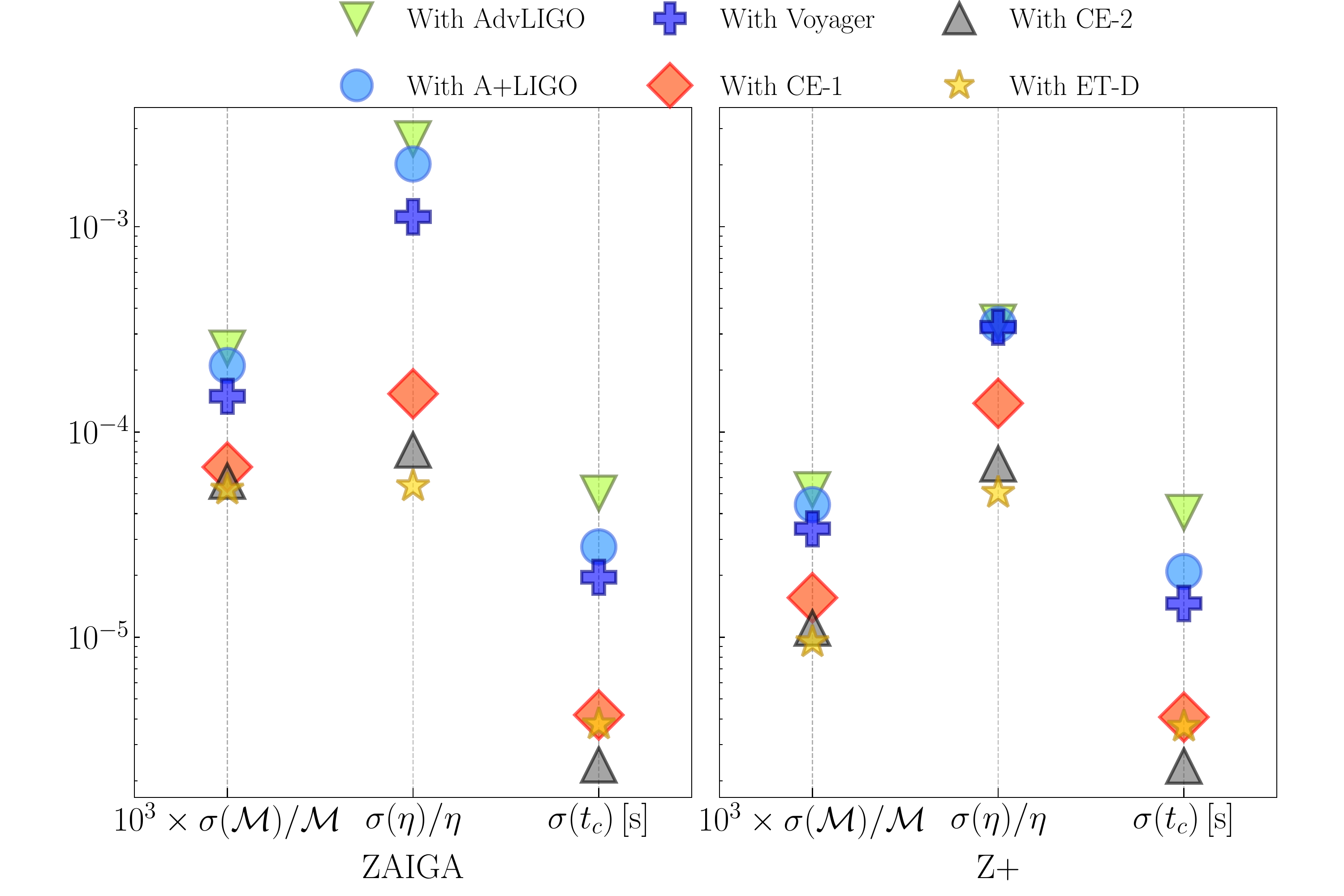}
    \caption{The median of the joint bounds on other parameters, $\mathcal{M}$, $\eta$,
    and $t_c$.}
    \label{fig:other_parameters}
\end{figure}

In~\cref{fig:individual_90CL}, we illustrate the constraints on $B$, namely
$\sigma(B)$, from individual ground-based GW detectors. It is evident that,
for ground-based laser-interferometer detectors, ET-D can provide the
tightest constraint on the parameter $B$, where $\sigma_{\rm com}(B)$ is
about $9 \times 10^{-10}$. It is due to its good sensitivity and its low
accessible frequency, starting from $1\,$Hz (see Table~\ref{tab:paras}). As
for atom-interferometer detectors, benefiting from their lower accessible
frequency than that of laser-interferometer detectors, the bounds on $B$
can reach $\sigma(B) \lesssim 10^{-8}$ in ZAIGA and $ \lesssim 10^{-9}$ in
Z+. However, the SNRs in these two detectors are rather low, namely
$\rho_{\rm mean} \sim 0.3$ for ZAIGA and $\rho_{\rm mean} \sim 3$ for Z+.
These low SNRs in ZAIGA and Z+ will prevent us from positively detecting
the BNS signals, and thus the bounds on $B$ from FIM are meaningless when
only the atom interferometers are considered~\cite{Vallisneri:2007ev}.

We expect that technically speaking, although with very low SNRs for
signals, ZAIGA/Z+ can record all the strain data for BNS mergers
whatsoever. These ``GW signals'' will be buried deep in noises. Late in the
frequency evolution of these signals, these BNSs will positively trigger
laser-interferometer detectors with enough SNRs. Then, we can confidently
assume that these GW signals are also recorded in atom-interferometer
detectors, though not being loud enough to trigger a detection. With this
assumption, we can combine data from atom interferometers and laser
interferometers. Such a combination is not useful to obtain a larger SNR
but will put tight constraints on negative PN terms, as they are so
sensitively dependent on the low-frequency data. Thus, a tighter constraint
on $B$ can be derived from the recorded ZAIGA/Z+ data with the help of
other detectors. To our knowledge such kind of combination is firstly
proposed in literature. The details certainly deserve further
investigation.

With the above guidelines in mind, in~\cref{fig:joint_90CL}, we focus on
the joint constraints from ZAIGA/Z+ together with the other six
laser-interferometer detectors. As we can see, with the contributions from
ZAIGA/Z+, the joint constraints on $B$ are improved significantly for laser
interferometers. Especially, for AdvLIGO, A+LIGO, and Voyager, the joint
constraints $\sigma(B)$ can be tightened to $\lesssim 1 \times 10^{-9}$
with ZAIGA. It is improved by more than an order of magnitude. It means
that, in the future joint observations, AdvLIGO, A+LIGO, Voyager, and CE
can help ZAIGA/Z+ to extract BNS parameters, and then ZAIGA/Z+ can help
them to constrain the dipole radiation in return. In addition, we also
notice that the bounds on dipole radiation from ZAIGA/Z+ alone are similar
to the results from the joint constraints when ZAIGA/Z+ is combined with
other laser-interferometer detectors. Although the constraints from
ZAIGA/Z+ alone in \cref{fig:individual_90CL} are derived from inappropriate
FIM calculation, they still show ZAIGA/Z+'s potential in testing dipole
radiation. In this sense, it is easy to understand that in the joint
analysis, ZAIGA/Z+ plays the key role in constraining the dipole radiation,
albeit with very low SNRs for BNS events. Of course, ZAIGA and Z+ are very
hard to properly constrain the dipole radiation as BNSs are hardly detected
in these detectors without the help of laser-interferometer detectors.
Therefore, it will be of great help to estimate the possible bounds on the
dipole radiation by considering the joint observation in the future. 
As we see from the figure, Z+ can help ET improve the
constraint on $B$ from $\sigma(B) \sim 2\times 10^{-9}$ to $\sigma(B) \sim
1 \times 10^{-10}$. Furthermore, in~\cref{fig:joint_90CL}, it is shown
that, compared to ground-based laser interferometers alone, the tighter bounds on $B$
can ultimately reach the levels of $|B| \lesssim 10^{-9}$ with ZAIGA and $|B|
\lesssim 10^{-10}$ with Z+ in the joint observations.

Now, we turn to the bounds on the other parameters, such as
$\mathcal{M}$, $\eta$, and $t_c$. The constraints from individual GW detectors
and joint observations including ZAIGA/Z+ are shown in~\cref{tab:population}
and~\cref{fig:other_parameters}, respectively. In addition to $B$, the bounds on
$\mathcal{M}$ and $\eta$ are improved with the help of ZAIGA/Z+, relative to the
bounds from laser-interferometers only. The joint constraints on $t_c$ are
similar to these from laser-interferometers alone. It indicates that ZAIGA/Z+ can
hardly improve the sky location in the joint observations.

\section{Discussions}
\label{sec:dis}

In this paper, we studied the projected bounds from near-future
ground-based GW detectors on dipole radiation using a popular
parametrization (\ref{eq:B}) by \citet{Barausse:2016eii}. Gravitational
dipole radiation could come from some alternative theories with additional
gravitational degrees of freedom in addition to $g_{\mu \nu}$ in GR. Our
work put the focus on BNS systems. It is relevant to some class of
alternative gravity theory, e.g. the Damour-Esposito-Far\'ese theory. We
used the modified GW waveforms augmented by dipole radiation in the ppE
framework~\cite{Yunes:2009ke, Tahura:2018zuq}. Meanwhile, we considered the
population property of BNS systems and simulated possible BNS systems to be
detected by ground-based laser-interferometer and atom-interferometer
detectors. After performing parameter estimation with the FIM technology,
we obtained the distribution of bounds on the dipole radiation. To
investigate the effects of atom-interferometer detectors (ZAIGA/Z+ in our
paper), we compared different scenarios in detail. The main conclusions are
summarized in the following.
\begin{enumerate}[(i)]
  \item For the atom-interferometer detectors ZAIGA and Z+, due to their
  limited sensitivity, it is hard to detect a distant BNS system with a
  high SNR. However, they are still useful in constraining the dipole
  radiation when combined with the confident detections from other
  ground-based laser-interferometer detectors. In that way, the constraints
  on dipole radiation can be tightened significantly in the future.
  \item In spite of low SNRs, the constraints on $B$ from ZAIGA/Z+ are
  tighter than most ground-based laser interferometer detectors. The
  constraints derived from the {\it inappropriate} FIM method for ZAIGA/Z+
  can be used as an order-of-magnitude estimate. According to the results
  of Scenario (II) in Sec.~\ref{sec:method}, where atom interferometers are
  combined with laser interferometers, ZAIGA/Z+ can help other ground-based
  detectors get tighter bounds.\footnote{It might be better to use other methods like the Markov-chain
  Monte Carlo (MCMC) method, see e.g., Ref.~\cite{Toubiana:2020cqv}, to confirm the constraints on
  $B$ derived from FIM by combining the low SNR waveforms in the atom-interferometer band with the waveforms in the laser-interferometer
  band. This is left for future work because the MCMC method is
  time-consuming.}
  \item Due to its low accessible frequency and high sensitivity, ET can
  provide a tight bound on $B$ alone, $\sigma(B) \sim 2 \times 10^{-9}$.
  Even so, the constraint can still be improved to the level of ${\cal O}
  (10^{-10})$ from its joint observation with Z+.
\end{enumerate}

As summarized above, the atom-interferometer detectors ZAIGA/Z+ may have
limited potential to detect a GW from realistic BNSs alone, but they can
complement other ground-based laser-interferometer detectors in testing the
gravitational dipole radiation from BNS mergers. This is largely due to the
low accessible frequency for atom interferometers. In the future, we expect
the constraints on dipole radiation can be tightened with different types
of ground-based GW detectors' complementary roles. The elements crucial to
the SEP, which are closely associated with the dipole radiation in
strong-field gravity, can be tested further. Meanwhile, as we have shown in
this work, probing dipole radiation with BNS systems can be an additional
science case for building atom-interferometer detectors like ZAIGA/Z+.

\acknowledgments 

We thank the  anonymous referee for constructive comments that improved the work.
We are grateful to Dong-Feng Gao, Huimei Wang, and Ming-Sheng Zhan for
helpful discussions. This work was supported by the National Natural
Science Foundation of China (11975027, 11991053, 11721303), the National SKA Program of China (2020SKA0120300),
the Young Elite Scientists Sponsorship Program by the China Association for
Science and Technology (2018QNRC001), and the Max Planck Partner Group
Program funded by the Max Planck Society. It was partially supported by the
Strategic Priority Research Program of the Chinese Academy of Sciences
through the Grant No. XDB23010200, and the High-performance Computing
Platform of Peking University.

\appendix

\section{Fisher Information Matrix}
\label{sec:app:FIM}

In estimation of the parameter set $\boldsymbol{\xi}$, when the SNR is
large, the probability $p(\Delta \boldsymbol{\xi})$ is characterized
by~\cite{Finn:1992wt,Finn:1992xs},
\begin{equation}
    \ln p(\Delta \boldsymbol{\xi}) \propto - \frac{1}{2} \Gamma_{ij} \, \Delta \xi_i \, \Delta \xi_j \,,
\end{equation}
where $\Delta \xi_i \equiv \xi_i - \hat{\xi}_i$, with the
maximum-likelihood estimate $\hat{\xi}_i$ determined in the
matched-filtering method. The variance-covariance matrix for estimating the
parameters $\boldsymbol{\xi}$ with a network of detectors is derived by
inversing the network FIM, which is obtained by summing the individual
Fisher matrices~\cite{Finn:1992wt,Will:1994fb},
\begin{equation}
    \Gamma_{i j}=\sum_{\alpha} \Gamma_{i j}^{\alpha} \,.
\end{equation}

To calculate the FIM, we need to compute partial derivatives of
$\widetilde{h}(f)$ with respect to different parameters. For GW amplitude
$\calA$, the time at coalescence $t_c$, the phase at coalescence $\phi_c$, and the dipole parameter $B$, we use the analytical results,
\begin{align}
    \frac{\partial \htilde(f)}{\partial \ln \calA} &= 
        \htilde(f)\,, \\
    \frac{\partial \htilde(f)}{\partial t_c} &= \ii 2 \pi f \,  \htilde(f) \,, \\
    \frac{\partial \htilde(f)}{\partial \phi_c} &= -\ii \,
        \htilde(f)\,, \\    
    \frac{\partial \htilde(f)}{\partial B} &= 
        - \ii \frac{3}{224 \eta} (\pi M f)^{-7 / 3} \, \htilde(f) \,.
\end{align}
For the other parameters, including the aligned dimensionless spins, the
chirp mass, and the symmetric mass ratio, the partial derivatives are
obtained numerically by
\begin{equation}
    \frac{\partial \htilde(f)}{\partial \xi_{i}} \approx 
    \frac{\Delta \htilde(f)}{\Delta \xi_{i}} \equiv 
        \frac{\htilde\left(\xi_{i}+\Delta \xi_{i};
        f\right)-\htilde\left(\xi_{i}; f\right)} {\Delta \xi_{i}} \,.
\end{equation}


%

\end{document}